\documentclass[11pt,aps,prc,superscriptaddress,showpacs,floatfix]{revtex4-1}
\usepackage[usenames, dvipsnames,dvipdfmx]{color} 
\usepackage{graphicx}
\usepackage{amssymb}
\usepackage{bm}
\usepackage{revsymb}
\usepackage{amsmath}
\usepackage{dcolumn}
\usepackage[normalem]{ulem}
\usepackage{caption}
\usepackage{subcaption}
\usepackage{ragged2e} 
\DeclareCaptionJustification{justified}{\justifying}
\usepackage{epsfig}
\usepackage{mathrsfs}
\usepackage{amsmath,amssymb}
\usepackage{natbib}
\usepackage{multirow}
\usepackage{booktabs, makecell, multirow}
\usepackage{eqparbox}

\begin{document}
\draft
\title{Neutrino self-interaction and MSW effects on the supernova neutrino-process}
\author{Heamin Ko}
\author{Myung-Ki Cheoun}
\thanks{\textrm{e-mail:}cheoun@ssu.ac.kr}
\author{Eunja Ha}
\affiliation{Department of Physics and OMEG institute, Soongsil University, Seoul 07040, Korea}
\author{Motohiko Kusakabe}
\affiliation{School of Physics and International Research Center for Big-Bang Cosmology and Element Genesis, Beihang University, Beijing 100083, China}
\author{Takehito Hayakawa}
\affiliation{National Institutes for Quantum and Radiological Science and Technology, 2-4 Shirakata, Tokai, Naka, Ibaraki 319-1106, Japan}
\author{Hirokazu Sasaki}
\affiliation{The University of Tokyo, Bunkyo-ku, Tokyo 113-0033, Japan \\ and National Astronomical Observatory of Japan, Mitaka, Tokyo 181-8588, Japan}
\author{Toshitaka Kajino}
\affiliation{School of Physics and International Research Center for Big-Bang Cosmology and Element Genesis, Beihang University, Beijing 100083, China}
\affiliation{The University of Tokyo, Bunkyo-ku, Tokyo 113-0033, Japan \\ and National Astronomical Observatory of Japan, Mitaka, Tokyo 181-8588, Japan}
\author{Masa-aki Hashimoto}
\affiliation{Kyushu University, Hakozaki, Fukuoka 812-8581, Japan}
\author{Masaomi Ono}
\affiliation{RIKEN, 2-1 Hirosawa, Wako-shi, Saitama 351-0198, Japan}
\author{Mark D. Usang}
\affiliation{Tokyo Institute of Technology, 2-12-1 Ookayama, Meguro, Tokyo 113-0033, Japan}

\author{Satoshi Chiba}
\affiliation{Tokyo Institute of Technology, 2-12-1 Ookayama, Meguro, Tokyo 113-0033, Japan}

\author{Ko Nakamura}
\affiliation{Fukuoka University, 8-19-1 Nanakuma, Jonan-ku, Fukuoka 814-0180, Japan}

\author{Alexey Tolstov}
\affiliation{Kavli Institute for the Physics and Mathematics of the Universe (WPI), The University of Tokyo, 5-1-5 Kashiwanoha, Kashiwa, Chiba 277-8583, Japan}

\author{Ken'ichi Nomoto}
\affiliation{Kavli Institute for the Physics and Mathematics of the Universe (WPI), The University of Tokyo, 5-1-5 Kashiwanoha, Kashiwa, Chiba 277-8583, Japan}
\author{Toshihiko Kawano}
\affiliation{Theoretical Division, Los Alamos National Laboratory, Los Alamos, NM 87545, USA}

\author{Grant J. Mathews}
\affiliation{Department of Physics, Center for Astrophysics, University of Notre Dame, Notre Dame, IN 46556, USA}

\date{\today}

\begin{abstract}
We calculate the abundances of $^{7}$Li, $^{11}$B, $^{92}$Nb, $^{98}$Tc, $^{138}$La, and $^{180}$Ta produced by neutrino $(\nu)$ induced reactions in a core-collapse supernova explosion.
We consider the modification by  $\nu$ self-interaction ($\nu$-SI) near the neutrinosphere and  the Mikheyev-Smirnov-Wolfenstein effect in outer layers for time-dependent neutrino energy spectra.
Abundances of $^{7}$Li and heavy isotopes  $^{92}$Nb, $^{98}$Tc and $^{138}$La are reduced by a factor of 1.5$\--$2.0 by the $\nu$-SI.
In contrast, $^{11}$B is relatively insensitive to the $\nu$-SI.
We find that the abundance ratio of heavy to light nucleus, $^{138}$La/$^{11}$B, is sensitive to the neutrino mass hierarchy, and the normal mass hierarchy is more likely to be consistent with the solar abundances.
\end{abstract}

\maketitle
\raggedbottom
The neutrino ($\nu$)-process is the nucleosynthesis mechanism induced by the neutrinos produced in core-collapse supernova (CCSN) explosions \cite{Woos90,Heger:2003mm}.  It is a unique nucleosynthesis process that only affects the abundances of some rare nuclei, such as $^{7}$Li and $^{11}$B \cite{Yoshi05,Yoshida06}, $^{19}$F \cite{Kobayashi:2011kf}, $^{92}$Nb \cite{Haya13} and $^{98}$Tc \cite{Haya17}, $^{138}$La, and $^{180}$Ta \cite{Heger:2003mm,Hayakawa:2010zza,Wu:2014kaa}.
A comparison of calculated $\nu$-process abundances  with observational abundances or meteoritic analyses can provide  valuable information on the associated $\nu$ physics and CCSN physics \cite{Heger:2003mm,Austin2011,Mathews12,Sieverding:2018rdt,Tamborra14, Abbar18, Glas19}. For example, recent progress of meteoritic analyses has revealed the ratios at the solar system formation, $^{92}$Nb/$^{93}$Nb $\simeq 10^{-5}$ \cite{Iizu16} and $^{98}$Tc/$^{98}$Ru $< 6 \times 10^{-5}$ \cite{Beck03}. This enables one to use both ratios as nuclear cosmochronometers for the duration from the last SN to the time of the solar-system formation \cite{Haya13,Haya17}.
The previous studies \cite{Heger:2003mm,Yoshi05,Wu:2014kaa} have also shown that the $\nu$-process isotopic abundances are sensitive to neutrino energy spectra, and consequently the $\nu$-process is a probe of the neutrino physics.

However, there still remain some ambiguities in treating the $\nu$ physics in CCSNe. One example is the $\nu$ mass hierarchy (MH), i.e. the normal hierarchy (NH) versus the inverted hierarchy (IH). The neutrino MH strongly affects the $\nu$-flux and the subsequently produced  $\nu$-process  abundances \cite{Yoshida06}. Another is the matter-enhanced $\nu$ oscillation, i.e. MSW effect, which  gives rise to additional $\nu$ mixing from that of free space around the bottom of the C/O-rich layer \cite{Yoshi05,Yoshida06}. The third important aspect is $\nu$ self-interaction ($\nu$-SI) arising from non-linear $\nu$-$\nu$ scattering \cite{Raff93,Volpe14,Chakraborty16,Sawyer16,Dasgupta17,Tamborra17}.
This is usually negligible, but near to the neutrinosphere the $\nu$-density  approaches  $\sim 10^{32} {\rm cm}^{-3}$  \cite{Pehl11}. This density is large enough that the $\nu$-SI should be taken into account for the estimation of the $\nu$-flux.
The previous study \cite{Wu:2014kaa} systematically calculated the $\nu$-process and ${\nu}p$-process considering the $\nu$-SI and MSW effects;
it was found that the abundances of $^{138}$La and $^{180}$Ta are enhanced by the $\nu$-SI  effect but the ${\nu}p$-process is not sensitive to this effect.
Because each $\nu$-process isotope is predominantly produced by one or two $\nu$-induced reactions \cite{Haya17}, its abundance is more sensitive to neutrino energy spectra rather than the other nucleosyntheses.
Recently, the $\nu$-SI effect to the ${\nu}p$-process was studied by including the multiangle three-flavor mixing \cite{Sasa17}.
In this paper, we report on the systematic investigation of the $\nu$-process by taking into account both the $\nu$-SI effect calculated from Ref. \cite{Sasa17} and the MSW effect.
We also discuss the MH dependence on heavy-to-light $\nu$-process isotopes.

All of the modifications due to the $\nu$-SI and the matter effect in the propagating $\nu$-flux can be  taken into account by solving the following evolution equation for the $\nu ({\bar \nu})$-density matrix \cite{Raff93,Volpe14}
\begin{eqnarray} \label{Boltzmann}
  i~{\dot \rho}_{\bold p} ({\dot {\bar \rho}}_{\bold p}) &=& + (-) { 1 \over {2E}} [ M^2, \rho_{\bold p} ({\bar \rho}_{\bold p})] + \sqrt{2} G_F [ L, \rho_{\bold p} ({\bar \rho}_{\bold p}) ] \\
  &&+ \sqrt{2} G_F \Sigma_{\bold q} ( 1 - \cos \theta_{{\bold p}{\bold q}} ) [ ( \rho_{\bold q} - {\bar \rho}_{\bold q}), \rho_{\bold p} ( {\bar \rho}_{\bold p}) ]. \nonumber
\end{eqnarray}
Here, $M$ is the $\nu$ mass-matrix including the vacuum oscillations, while ${\bold p}$ and ${\bold q}$ are the momenta of the propagating and background neutrinos. The $\nu$-density matrix $\rho$ and the charged lepton number density matrix $L$ are given by $\rho_{\alpha\beta} =\left\langle \nu_{\alpha} | \nu_{\beta}\right\rangle = \sum_{\gamma = e, \mu, \tau} \left\langle \nu_{\alpha} | \nu_\gamma (t) \right\rangle \left\langle \nu_\gamma (t) |\nu_{\beta}\right\rangle $ and $L_{\alpha\beta} =(N_\alpha -N_{\bar{\alpha}}) \delta_{\alpha \beta}$ with $\nu$ flavors $\alpha$ and $\beta$. $N_{\alpha}$ denotes the lepton density and $\delta_{\alpha\beta}$ is the Kronecker delta.
The first and second terms on the r.h.s. of Eq.\ (\ref{Boltzmann}) describe  $\nu$ oscillations in vacuum and matter, respectively. The electron density is calculated with a constant electron fraction w.r.t.\ the baryon density given by a fit \cite{Fogli:2003dw} to a shock-propagating model. The muon and tau densities are assumed to be negligible in this work. The $\nu$-SI is taken into account in the third term.

The evolution of the $\nu$-flux by the $\nu$-SI is achieved by solving Eq.\ (\ref{Boltzmann}) for the $\nu$ distribution function, $\ f(r;{\epsilon}_{\nu}, T_{\nu_{\alpha}} (t)) = f_{Fermi~Dirac} ( {\epsilon}_{\nu}, T_{\nu_{\alpha}} (t)) \langle \rho (r;{\epsilon}_{\nu}) \rangle $, which is normalized with the angle-averaged $\nu$-density matrix $\langle \rho ({r;\epsilon}_{\nu}) \rangle$.
The differential $\nu$-flux is defined as follows
\begin{equation}\label{flux}
 \frac{d}{d\epsilon_{\nu}}\phi_{\nu_\alpha}(t,r;\epsilon_{\nu},T_{\nu_{\alpha}})
  = \frac{{\cal L}_{\nu_{\alpha}}(t)}{4 \pi r^2} \frac{\epsilon_{\nu}^2}
 {\langle \epsilon_{\nu} \rangle} {f(r;{\epsilon}_{\nu},T_{\nu_\alpha} (t))}~,
\end{equation}
where ${\cal L}_{\nu_\alpha}(t)$ is the luminosity of $\nu_{\alpha}$.
We adopt the neutrino luminosity evolution based upon the 20 M$_\odot$ progenitor numerical CCSN simulations summarized in Ref. \cite{OConnor:2018sti}, where it was demonstrated that a variety of independent   numerical simulations produce nearly identical neutrino spectra and time evolution. Values for the ${\cal L}_{\nu_\alpha}$ and the averaged energy are deduced at $t = $ 5, 100, 200, 300 and 500 ms \cite{OConnor:2018sti} (see, Table I).
We do not consider the early neutrino burst $t < 50$ ms and assume an exponential decay in the $\nu$-luminosity after 500 ms.
Note that the ${\cal L}_{\nu_x} $ becomes weaker than the other luminosities with time while $ \langle E_{\nu_x} \rangle$ attains almost the same effective energy.

\begin{table}[h!]
\captionsetup{justification=centering,singlelinecheck=false}
    \caption{Time evolution of the luminosity ${\cal L}_{\nu_{\alpha}}$ and the effective energy $\langle E_{\nu_{\alpha}} \rangle $ from the neutrino transport models in Ref.\cite{OConnor:2018sti}. Here $\nu_x = \nu_\mu, \nu_\tau, {\bar \nu}_\mu$ and ${\bar \nu}_\tau$.}
    \begin{tabular}{c | c c c | c c c }
      \hline
      \hline
      time &  ~${\cal L}_{\nu_e}$  & ${\cal L}_{\bar{\nu}_e}$ & ${\cal L}_{\nu_x}$ &
      ~~$\langle E_{\nu_e} \rangle $ & $ \langle E_{\bar{\nu}_e} \rangle $ & $ \langle E_{\nu_x} \rangle $  \\
      \hline
      [ms]  & ~~~  &   [$ 10^{52}$ erg/s] & ~~  &
       &  [MeV] &    \\
      \hline
     50  ~&~ 6.5  & 6.0 & 3.6 ~&~  9.3 	& 12.2	& 16.5	\\
     \hline
     100   ~&~ 7.2  & 7.2  & 3.6 ~&~  10.5	& 13.3	& 16.5	\\
     \hline
     200   ~&~ 6.5  & 6.5  & 2.7 ~&~  13.3	& 15.5	& 16.5	 \\
     \hline
     300   ~&~ 4.3  & 4.3  & 1.7 ~&~  14.2	& 16.6	& 16.5	 \\
     \hline
     500   ~&~ 4.0  & 4.0  & 1.3 ~&~  16.0	& 18.5	& 16.5	\\
     \hline
    \end{tabular}
\label{Lumino}
\end{table}

\begin{figure}[h]
\begin{center}
{\includegraphics[width=16.8cm]{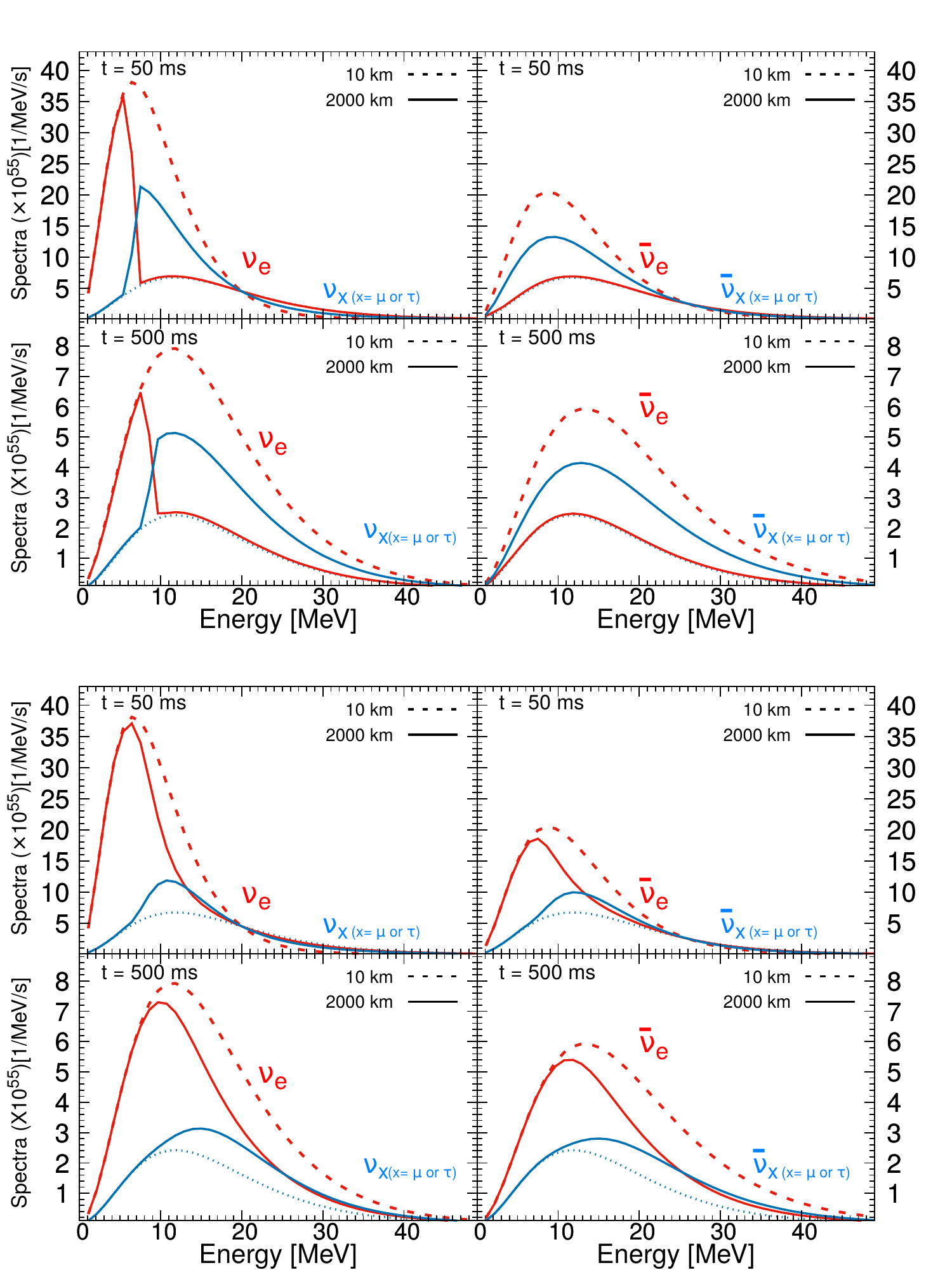}}
\caption{ \small{Differential $\nu$-flux $\phi^{\prime} (\equiv d \phi / d \epsilon)$ deduced from the neutrino luminosity ${\cal L}_{\nu_{\alpha}}$ \cite{OConnor:2018sti} and their modifications by the $\nu$-SI. The upper (lower) panels are for the IH (NH) case for $t$ = 50 and 500 ms. The left (right) panels are for $\nu$ (${\bar \nu}$). Dashed and solid lines show the initial flux at 10 km and the final flux at 2000 km after the $\nu$-SI, respectively.}}
\label{fig1}
\end{center}
\end{figure}

The neutrino and electron densities near the neutrinosphere play vital roles during the $\nu$-process in the SN environment. For instance, if the electron density is much higher than the $\nu$-density, it causes  suppression of the $\nu$-SI effect \cite{Chakraborty:2011nf}. However, as the shock wave propagates, the electron density decreases, so that the flavor change by the $\nu$-SI becomes significant in the outer region \cite{Duan15}.
Once the $\nu$-flux is changed by the $\nu$-SI, the flux distributions retain their shapes until they undergo the MSW effect.
The baryon matter density in the inner region depends upon the SN model employed. For our purposes, however,  it is adequate to adopt the phenomenological model of Fogli et al. \cite{Fogli:2003dw} (FLMM). Hence, we take a density profile for the inner region approximated as a power law and assume that it remains valid for $t \le 1 s$.
Neutrinos calculated by the FLMN density profile propagate from $r =10$ km to 2000 km with the $\nu$-SI, beyond which no changes by the $\nu$-SI occur.

To obtain the temperature and density profiles from the shock propagation we utilize  the pre-supernova (pre-SN) model developed for SN1987A \cite{Shigeyama88,Kikuchi15,Blinnikov00}.
The adopted hydrodynamics model of SN is constructed with the initial condition \cite{Blinnikov00} selected to reproduce the light curve of SN 1987A.
This model is for  a 16.2 $M_\odot$ progenitor with  a 6 $M_\odot$ He core and a metallicity of $Z = Z_{\odot} /4$, and the stellar evolution and nucleosynthesis have been updated with the method in Ref.~\cite{Kikuchi15}.
The weak $s$-process utilized the (n,$\gamma$) reaction data \cite{Kawano} for the A=100 mass region to obtain pre-SN abundances \cite{Haya17}.
For the $\nu$-process, we adopt a nuclear reaction network \cite{Moto19} and employ the previous numerical results \cite{Yoshida:2008zb} for $\nu$-nucleus reaction cross sections of the light nuclei.
These are calculated in a few-body model for the $^4$He reaction and in a shell-model for $^{12}$C. For the heavy nuclei, $\nu$-induced reactions are calculated in the quasi-particle random phase approximation through many multipole transitions dominated by the Gamow-Teller transition \cite{Cheoun:2011hj,Ch10-1}. Neutrino reaction rates in the SN explosion are calculated as follows
%
\begin{eqnarray}
\lambda_{\nu_\alpha}(t,r)
&=&    \int_{0}^{\infty} \sum_{\beta=e,\mu,\tau} {d \phi_{\nu_{ \beta}} \over {d \epsilon_{\nu}}}(t-r/c,r=2000~{\rm km} ; \epsilon_{\nu_\beta})  \ \nonumber \\
  &&~~~~\times P_{\nu_\beta \nu_\alpha} (r;\epsilon_{\nu})\ Br(\epsilon_{\nu}) \ \sigma_{\nu_\alpha}(\epsilon_{\nu})\ d\epsilon_{\nu}~.
\end{eqnarray}
Here the $\nu$ reaction cross section, $\sigma_{\nu_\alpha} $, is multiplied by the branching ratio, $Br (\epsilon_{\nu})$, of the excited states calculated using  the statistical method \cite{Iwamoto16}. The flavor transition probability, $P_{\nu_\beta \nu_\alpha}$, includes the $\nu$ oscillations in matter based upon the mixing parameters \cite{Agashe:2014kda}.

\begin{figure}
\begin{center}
\includegraphics[width=8.6cm]{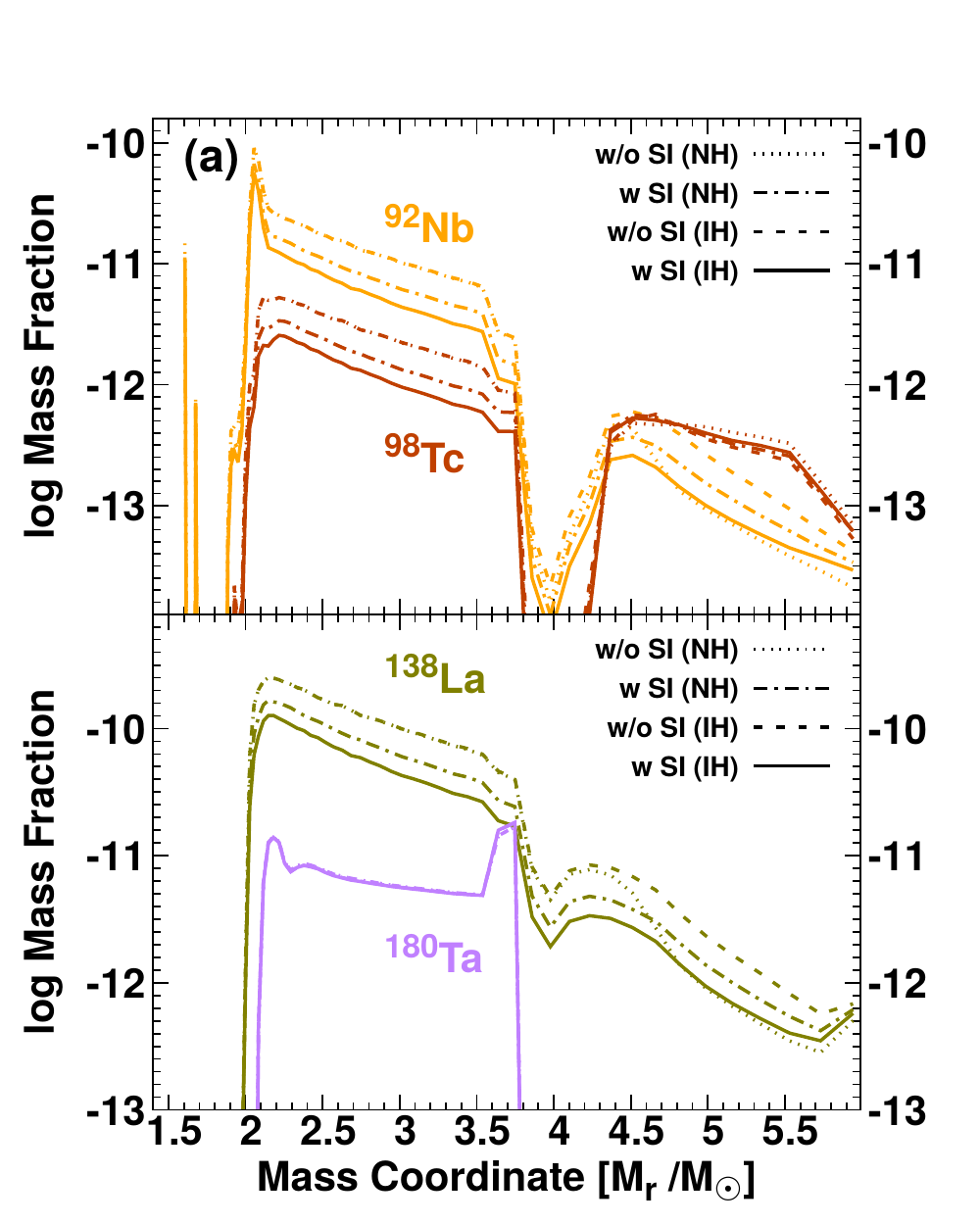}
\includegraphics[width=8.8cm]{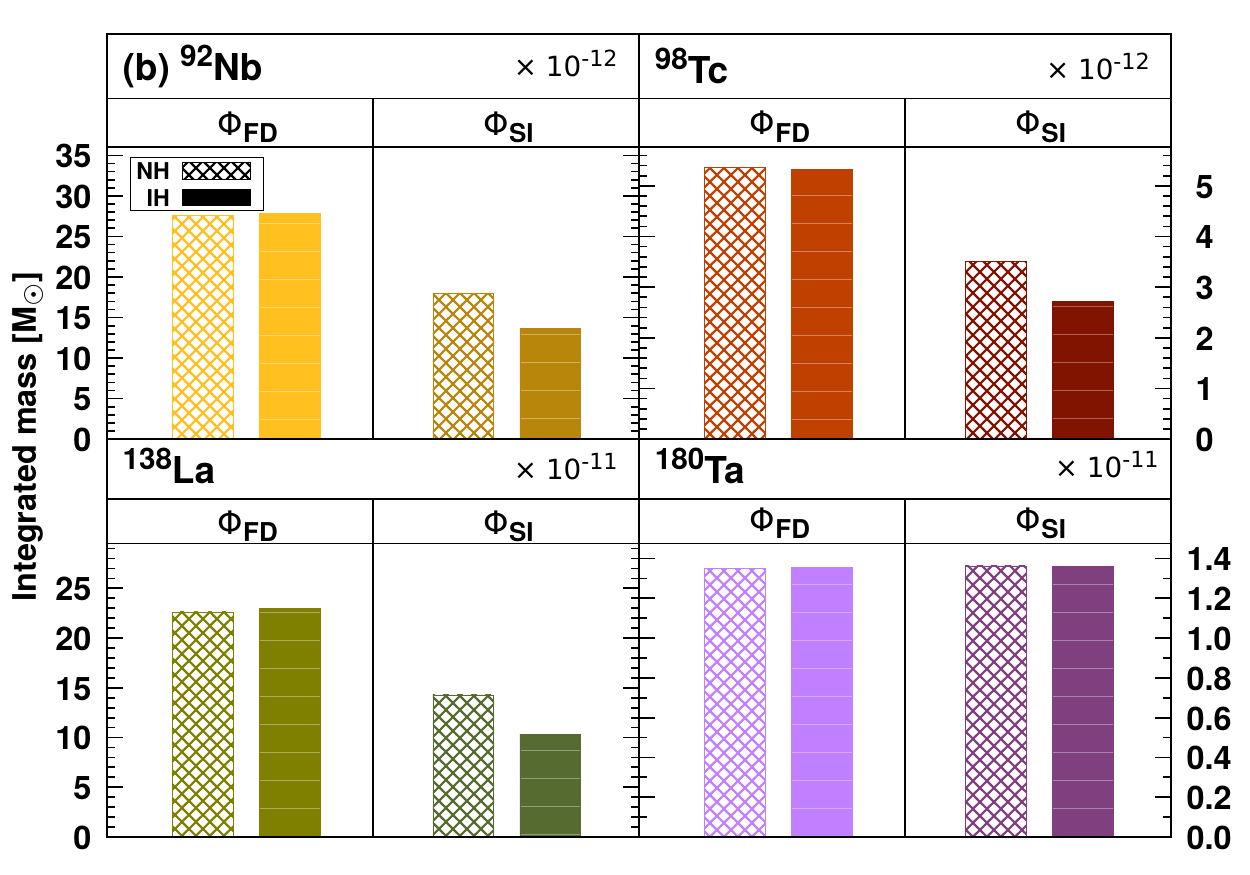}
\caption { \small{Mass fractions (a) and integrated masses (b) of $^{92}$Nb, $^{98}$Tc, $^{138}$La and $^{180}$Ta abundances in the NH and IH schemes. We show four different cases of w/o SI (NH) (dotted), w/ SI (NH) (dash-dotted), w/o SI(IH) (dashed) and w/ SI(IH) (solid).}}
\label{fig2}
\end{center}
\end{figure}

Numerical results of Eq.~(\ref{flux}) are presented in Fig.~\ref{fig1}.
They show how the $\nu$-flux emitted from the neutrinosphere is modified by the $\nu$-SI.
In the IH scheme, the $\nu_e$ ($\nu_{x= \mu, \tau}$) flux at 2000 km at $t = 50$ ms is lower (higher) than the original flux from the surface of the neutrinosphere in the energy range from about 6$\--$20 MeV.
The situation is again reversed in the higher energy region above the point of equal flux for
the three flavors.
However, at $t = 500$ ms, the point of equal flux becomes higher because of the higher $\langle E_{\nu_e} \rangle$.
As a result, the swapping for $\nu_e$ and $\nu_x$ occurs in a wide energy region above 8 MeV.
For anti-neutrinos, the swapping also occurs in a wider energy region (see, right panels in Fig.~\ref{fig1}).
In the NH scheme, these trends are also observed, but the result for the NH exhibits that the changes of spectra by the $\nu$-SI become weaker.
At t = 50 ms, the $\nu_e$ flux at 2000 km in NH scheme is higher than that in the IH scheme.
The present result shows that even if the average energies of $\nu_e$ and $\nu_x$ are identical, in the case that the luminosities of $\nu_e$ and $\nu_x$ are different the $\nu$-SI modifies their energy spectra and the final energy spectra depend on the MH.

Figure \ref{fig2} (a) shows the mass-fractions of $^{92}$Nb, $^{98}$Tc, $^{138}$La and $^{180}$Ta with and without the $\nu$-SI in each MH scheme. Abundances of $^{92}$Nb, $^{98}$Tc, and $^{138}$La decrease with increasing $M_r$ except for those in the valleys. This trend stems from the neutrino-induced reaction rate which is proportional to the neutrino flux which scales as $r^{-2}$. A valley around the $M_r$ $\sim$ 4 $M_\odot$ region results from strong destruction via the ($n$,$\gamma$) reactions behind the shock heating.
Another valley in the region of $M_r$ $<$ 2.0 $M_\odot$ comes from the photodisintegration of the pre-SN elements.
Note that the insensitivity of the $^{180}$Ta production to the $\nu$-SI comes from the fact that most of the $^{180}$Ta is not produced via the $\nu$-process in the present model.
Because most of the heavy nuclei are produced mainly inside the MSW region,
their abundances depend strongly on the $\nu$-SI.
We stress that the $\nu$-SI effect decreases the $^{92}$Nb, $^{98}$Tc and $^{138}$La abundances by a factor of 1.5$\--$2.0 and each final abundance in the NH scheme is larger than that in the IH scheme by about 20$\--$30\%.
These features are explicitly illustrated by the integrated masses in Fig. \ref{fig2} (b). This can be understood by the contribution of $\nu_e$ for the production.
These heavy nuclides are predominantly synthesized by charged current (CC) reactions with $\nu_e$ on pre-exiting nuclides such as the $^{138}$Ba($\nu_e$,~e$^-$)$^{138}$La reaction and its fraction by $\nu_e$ is as high as 70$\--$90 \% \cite{Heger:2003mm, Haya17}.
Thus, the decreased abundances by the $\nu$-SI can be attributed to the decrease of the $\nu_e$-flux.
Even if the average energies of $\nu_e$ and $\nu_x$ were nearly identical,
when the luminosity of $\nu_e$ is higher than that of $\nu_x$ the number of $\nu_e$ is decreased by the $\nu$-SI
and hence the $\nu$-process abundances are also decreased.

\begin{figure}[h]
\begin{center}
{\includegraphics[width=8.6cm]{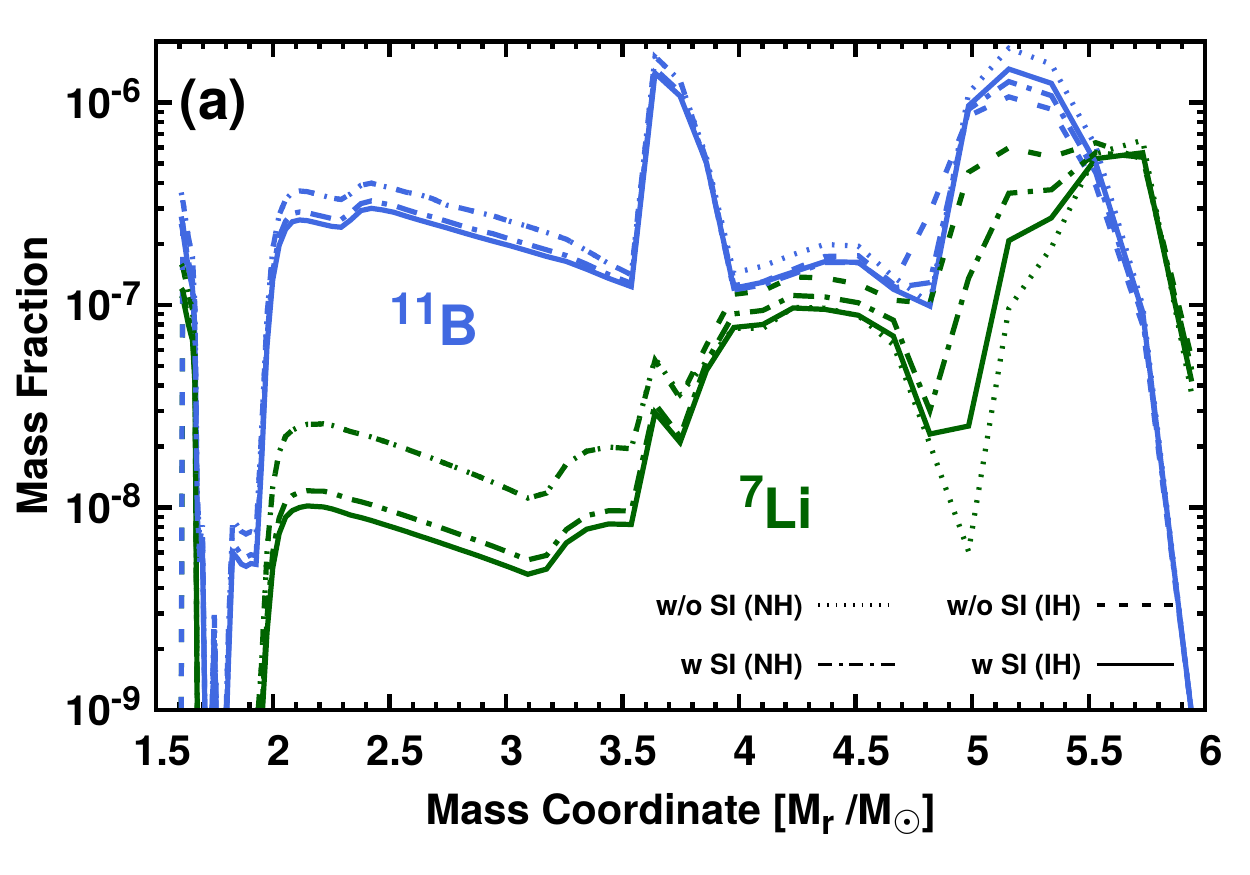}}
{\includegraphics[width=8.8cm]{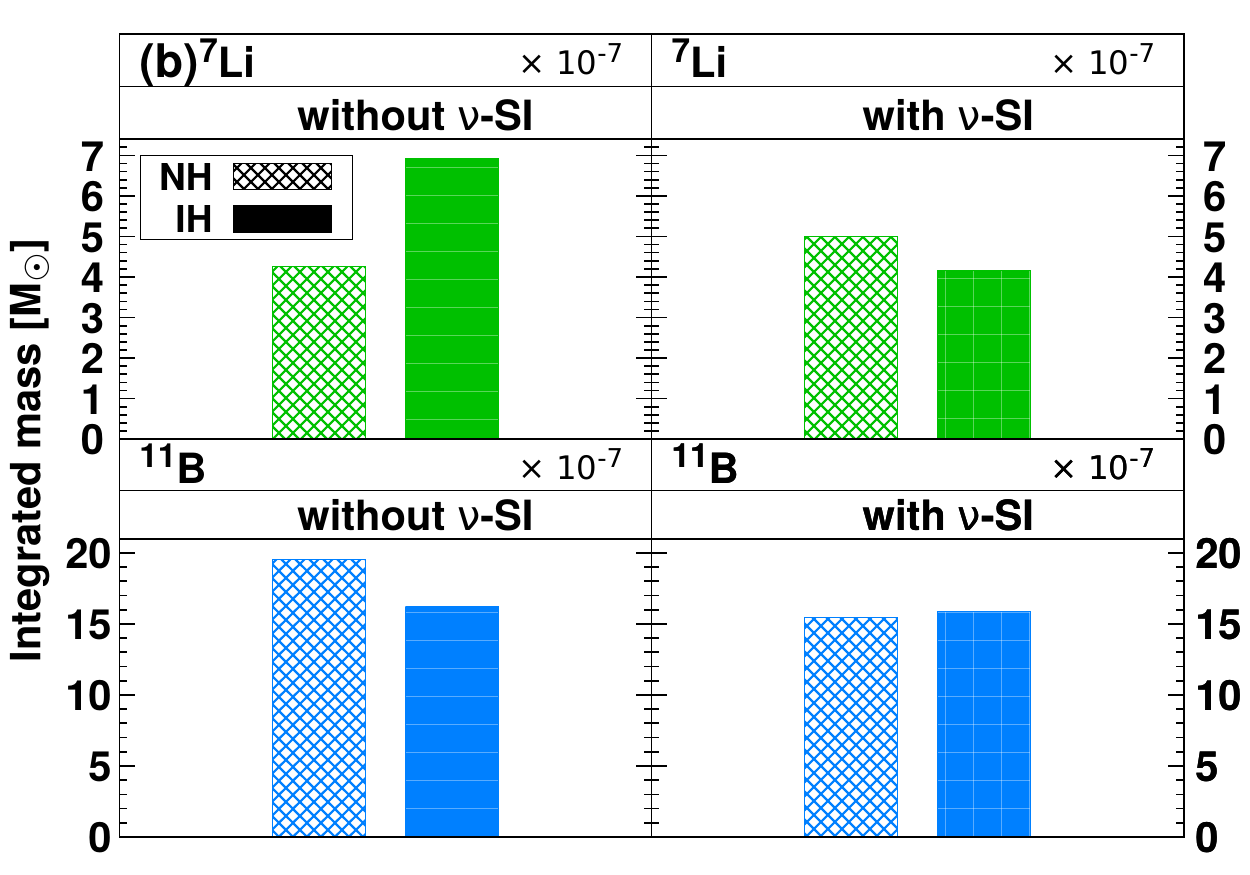}}
\caption{\small{Same as Fig.\ \ref{fig2}, but for $^{7}$Li and $^{11}$B. Abundances are plotted at 1 yr after the SN explosion. All results include the MSW effect.}}
\label{fig3}
\end{center}
\end{figure}

Figure \ref{fig3} (a) shows the abundances of the light nuclei, $^{7}$Li and $^{11}$B, including both the $\nu$-SI and the MSW effect, and their integrated masses are presented in Fig. \ref{fig3} (b). 
The main production regions are the outer region of the MSW layer, 4.7$\--$6.0 $M_\odot$.
The total abundance of $^{7}$Li is much decreased by the $\nu$-SI in the IH scheme, whereas in the NH scheme the $^{7}$Li abundance is slightly increased.
$^{7}$Li is produced from $^4$He by the $\nu_e$ and $\bar{\nu_e}$ via CC reactions as well as neutral current (NC) reactions \cite{Yoshida:2008zb}.
The cross sections of the two CC reactions are larger than those of the NC reactions by a factor of 2--3, and the cross section of CC reactions with $\nu_e$ is slightly larger than that with $\bar{\nu_e}$ \cite{Moto19,Yoshida:2008zb}.
As a result, the $^{7}$Li abundance is sensitive to $\nu_e$ flux and the $\nu$-SI effect on $^{7}$Li is similar to that on the heavy isotopes.
Although $^{11}$B is also generated by CC reactions with $\nu_e$ and $\bar{\nu_e}$ on $^{12}$C in addition to NC reactions, these three reactions have contributions of the same order of magnitude \cite{Yoshida:2008zb}.
Thus, $^{11}$B production is relatively insensitive to $\nu$-SI and its abundance decreases by only 5$\--$10\%. In addition, the difference between the IH and NH is only a few \%.
The previous study suggested that the abundance ratio $^{7}$Li/$^{11}$B is sensitive to the MH \cite{Yoshida06}.
The $^{7}$Li/$^{11}$B ratio is changed by the $\nu$-SI effect from 0.67 to 0.41 in the IH scheme, and from 0.34 to 0.51 in the NH scheme.
The $^{7}$Li/$^{11}$B ratio in the NH scheme is larger than that in IH by about 25\% in the present model.

Here we discuss the yield ratio of $^{138}$La and $^{11}$B. The previous study on the $\nu$-process without considering both the $\nu$-SI and the MSW effects \cite{Heger:2003mm} concluded that enough $^{138}$La is produced by the $\nu$-process, while the $^{11}$B is overproduced. The present result shows that the $^{138}$La abundance is decreased by a factor of about 2, whereas the $^{11}$B abundance is nearly indifferent to the $\nu$-SI.
A ratio of PF($^{138}$La) to PF($^{11}$B) defined as PF[A] = $X_A/X_{A\odot}$ with $X_A$ the mass fraction of A is changed by the $\nu$-SI.
The PF($^{138}$La)/PF($^{11}$B) ratio is approximately 0.26 and 0.18 for the NH and IH, respectively; the ratio in the NH scheme is larger than that in the IH scheme by a factor of about 1.4.
This large difference originates from the fact that $^{138}$La is predominantly produced by $\nu_e$ but $^{11}$B production is insensitive to $\nu$-SI  as discussed above.
$^{138}$La is considered to be produced predominantly by the SN $\nu$-process, whereas $^{11}$B is also produced by cosmic rays and a study \cite{Pra12} estimates that about 30\% of the $^{11}$B solar abundance originates from the $\nu$-process.
Therefore, the ratio of 0.26 in the NH is more likely to be consistent with the solar ratio of 0.3.
This trend originates from the fact that the abundance change by the $\nu$-SI in the IH scheme is stronger than that in the HM scheme.
After the $\nu$-SI effect, the $\nu_e$ flux for the NH scheme in a energy range of 10$\--$20 MeV, which is the effective energy region for $^{138}$La production, is higher than that for the IH scheme by a factor of 2$\--$3 (see Fig.~\ref{fig1}).
As discussed previously, $^{138}$La production depends strongly on $\nu_e$ flux.
Thus, if the initial neutrino energy spectra are changed from that assumed here, the trend that the PF($^{138}$La)/PF($^{11}$B) ratio after  the $\nu$-SI effect in the NH scheme is higher than that in the IH scheme is expected to be preserved.

Finally, we note that the recent three-dimensional hydrodynamical SN simulations  predicted asymmetric radiations of $\nu_e$ and $\bar{\nu_e}$ \cite{Tamborra14} and that the following studies taking the neutrino angular distribution into account suggest that if the angular distributions of $\nu_e$ and $\bar{\nu_e}$ are different the fast neutrino flavor transformation by crossing of $\nu_e$ and $\bar{\nu_e}$ occurs \cite{Chakraborty16,Sawyer16,Dasgupta17,Tamborra17}.
In this case, the energy swapping may occur in the earlier time and is affected by the larger different luminosities between $\nu_e$ and $\nu_x$. The hypothetical sterile neutrino may also cause fast neutrino flavor changes \cite{Jang19}.
This may enhance the MH dependence for $\nu$-process abundances.
However, the detailed calculation for more precise evaluation is beyond of the present scope.

In conclusion, we have included the effects of both the $\nu$-SI and MSW mixing on the $\nu$-process in CCSN explosions by adopting numerical results for the time-dependent $\nu$-luminosity.
Even if the average temperatures of neutrino flavors are almost same, when the luminosities of neutrino species are different the $\nu$-SI affects the $\nu$-process abundances.
Abundances of heavy $\nu$-isotopes and $^7$Li are reduced by a factor of 1.5$\--$2, whereas $^{11}$B is decreased only by 5$\--$10\%.
The reduction of the $\nu$-isotopic abundances can be systematically understood by the reduction of the $\nu_e$ flux by the $\nu$-SI.
The contribution of CC reactions with $\nu_e$ for production of $^{7}$Li and heavy $\nu$-process isotopes is relatively large, whereas for $^{11}$B the contributions of $\bar{\nu_e}$ and other neutrinos are of the same order as $\nu_e$.
Abundance ratios of heavy to light $\nu$-process isotopes such as $^{138}$La/$^{11}$B turn out to be more sensitive to the MH, and the present result comparing to the solar abundances shows that the NH scheme is favored.


\end{document}